**Title:**
Nano-scale imaging of the full strain tensor of specific dislocations extracted from a bulk sample

**Authors:**
Felix Hofmann[1,*], Nicholas W. Phillips[1], Suchandrima Das[1], Phani Karamched[2], Gareth M. Hughes[2], James O. Douglas[2], Wonsuk Cha[3], Wenjun Liu[3]

[1] Department of Engineering Science, University of Oxford, Parks Road, Oxford, OX1 3PJ, UK
[2] Department of Materials, University of Oxford, Parks Road, Oxford, OX1 3PH, UK
[3] Advance Photon Source, Argonne National Lab, 9700 S. Cass Avenue, Lemont, IL, USA

[*] felix.hofmann@eng.ox.ac.uk

**Abstract:**
Lattice defects play a key role in determining the properties of crystalline materials. Probing the 3D lattice strains that govern their interactions remains a challenge. Bragg Coherent Diffraction Imaging (BCDI) allows strain to be measured with nano-scale 3D resolution. However, it is currently limited to materials that form micro-crystals. Here we introduce a new technique that allows the manufacture of BCDI samples from bulk materials. Using tungsten as an example, we show that focussed ion beam (FIB) machining can be used to extract, from macroscopic crystals, micron-sized BCDI samples containing specific pre-selected defects. To interpret the experimental data, we develop a new displacement-gradient-based analysis for multi-reflection BCDI. This allows accurate recovery of the full lattice strain tensor from samples containing multiple dislocations. These new capabilities open the door to BCDI as a microscopy tool for studying complex real-world materials.

**I. Introduction**
Lattice defects can dramatically alter the properties of crystalline materials. In metals, dislocations provide a low energy pathway for plastic deformation [1]. Their interactions with one another and with other micro-structural features control material strength [1–3]. All modern alloys rely on microstructural engineering to control dislocation behaviour and thereby enhance properties [4–7]. Crystal defects interact via the distortions (i.e. the strain fields) they cause in the crystal lattice [1]. Understanding these strain fields is essential for engineering defect properties to enhance material performance. It is also key to explaining why seemingly insignificant changes in defect environment can cause substantial changes in mechanical properties. For example, addition of trace amounts of hydrogen (which strongly interacts with dislocations) dramatically modifies the mechanical properties of steel, posing major challenges for petrochemical and nuclear industries [8–11].

Quantifying the lattice distortions associated with specific defects is challenging. High Resolution Transmission Electron Microscopy (HRTEM) and Scanning Transmission Electron Microscopy (STEM) allow dislocation strain fields to be measured with atomic resolution [12–14]. This has provided much needed validation for theoretical predictions of dislocation strain fields [15–20]. However, TEM strain measurements are only possible on dislocations that are straight, are oriented normal to the thin foil surface, and lie along specific zone axes. This precludes measurement of the strains associated with dislocation structures and junctions that control material strength. Furthermore, these techniques only measure the in-plane components of the lattice strain tensor. While this is sufficient to characterise edge dislocations, it does not allow the analysis of screw dislocations where the Burgers vector (and hence the most prominent displacements) are parallel to the dislocation line. Even in simple metals such as tungsten there is still intense debate about the structure of screw dislocations [21–24].

Bragg Coherent Diffraction Imaging (BCDI) has emerged as a promising technique for 3D characterisation of morphology and lattice strain in micro-crystals [25,26]. In BCDI, coherent X-ray diffraction patterns (CXDP) are measured from lattice reflections of a micro-crystal illuminated with a coherent X-ray beam. Using the approximate formalism of an effective electron density, the far-field CXDP can be interpreted as the Fourier transform of the effective electron density in the sample. Unfortunately, while the intensity of the diffracted wave field can be reliably recorded, the phase information is lost. Hence one cannot simply inverse Fourier transform the CXDP to find the electron density. Phase retrieval algorithms must first be used to recover the lost phase information [27]. The reconstructed electron density is complex-valued. Its amplitude, $\rho(\mathbf{r})$, provides information about the morphology of the scattering crystal domain, where $\mathbf{r}$ is the spatial coordinate. The spatially varying phase of a *hkl* reflection, $\emptyset_{hkl}(\mathbf{r})$, is linked to the atomic displacement field of the crystal lattice, $\mathbf{u}(\mathbf{r})$, by $\emptyset_{hkl}(\mathbf{r}) = \mathbf{q}_{hkl} \cdot \mathbf{u}(\mathbf{r})$, where $\mathbf{q}_{hkl}$ is the reflection Bragg vector [28]. As such, BCDI allows non-destructive probing of both crystal morphology and distortion of the crystal lattice along the Bragg vector with nano-scale spatial resolution [29].

In the past decade BCDI has evolved from a niche technique to a mainstream scientific tool. It has been applied to a multitude of challenging scientific questions, from understanding charge-discharge-induced strains in battery nano-crystals [30], to probing growth and dissolution of organo-mineral crystals [31,32], to imaging nano-scale (de)alloying [33,34], to monitoring *in-situ* catalysis [35,36] or probing radiation damage evolution in protein crystals [37,38], to name but a few examples. A key limitation of BCDI is that it requires crystallographically-isolated micro-crystal samples in the size range from ~ 100 nm to ~ 1 μm; sufficiently large to give a strong scattering signal, but small enough to match the coherence volume of the X-ray beam [27]. Only a small number of materials form crystals that fall into this size range, for example metal micro-crystals (Au [35,39–42], Pt [36]), ceramic nano-rods [43,44] or metal thin films with grain size in this range [45,46]. The vast majority of technologically important materials, however, do not form suitable micro-crystals with grain size in this range. As a result, most BCDI experiments have been confined to prototypical studies performed on model systems.

In electron microscopy, focussed ion beam (FIB) machining has become a mainstream tool for the manufacture of samples from bulk specimens [47,48]. It has the advantage of allowing the targeted extraction of specific microstructural features for investigation. For example it has been used to produce specimens containing pre-selected crack tips for TEM [49–51] or atom probe tomography (APT) [52,53]. FIB is also being used extensively to make miniaturised material test samples, such as micro-pillars [54,55] or micro-cantilevers [56–58], to selectively probe the mechanical properties of particular microstructural features. Within the X-ray community the use of FIB to machine samples for high resolution imaging studies is becoming more widespread. For example, it has been used to produce samples for nano-tomography of integrated circuits [59], complex alloys [60,61] and magnetic materials [62]. Surprisingly, the ability of FIB to shape material at the nano-scale has not thus far been exploited to machine BCDI samples from larger metallic material volumes. A reason for this may be the extensive damage produced by FIB, which ranges from lattice defects [63] to amorphisation [64], the formation of intermetallic phases [65] and even the nucleation of twin domains [66]. For imaging approaches that are not sensitive to lattice strain, this damage has relatively little impact. However, for BCDI, which has excellent sensitivity to lattice strain, the presence of FIB damage causes severe complications. BCDI measurement of FIB damage in gold showed that even a single FIB imaging scan leaves behind large strain fields [39]. More extensive FIB machining caused the formation of an extended dislocation network [39,40], and gave rise to lattice strains that extend up to hundreds of nano-metres beneath the FIB-damaged surface [67]. These large strain fields complicate convergence of BCDI reconstructions, and can obscure more subtle strain fields of interest, for example those associated with crystal defects. For FIB to serve as a useful tool for the manufacture of BCDI samples, approaches that mitigate the effects of FIB damage are needed.

Previous BCDI studies that observed phase signatures from dislocations concentrated on a single crystal reflection [30,31,45,46,68,69]. This provides only one lattice displacement component, and hence only one of the six lattice strain components. However, for a direct comparison with dislocation simulations, access to the full lattice strain tensor is vital [70,71]. By combining the phase (i.e. lattice displacement) from BCDI measurements of three or more reflections with linearly independent scattering vectors, the full lattice displacement field can be recovered [39,44]. The full 3D resolved lattice strain tensor can then be determined by numerical differentiation of the 3D displacement field [39,40,44,72]. This approach works well in crystals where lattice displacement varies without discontinuities, such that any phase wraps can be reliably unwrapped using existing approaches (e.g. [73]). However, crystal defects, such as dislocations, lead to discontinuities in the displacement field, since the lattice to once side of the defect is displaced by one atomic spacing [1,31]. When differentiating to recover lattice strain, this leads to incorrect, large strains across the discontinuity. To correct this, the periodicity of the crystal lattice must be accounted for, which is challenging to do in a general way. A new approach to computing lattice strain tensor from multi-reflection BCDI measurements of dislocation-containing samples is required.

Here we present a new technique for manufacturing BCDI strain microscopy samples from bulk materials using FIB machining. Our method makes it possible to first identify specific defects of interest in a bulk specimen, and then create a micron-sized sample containing these defects. Importantly, it provides a reliable approach for minimising FIB damage and associated spurious lattice strain fields. This overcomes a key hurdle of previous BCDI studies, which required materials that naturally form micron-sized crystals and then relied on luck to place a suitable defect within these crystals. To enable faithful reconstruction of the full lattice strain tensor in dislocation-containing samples, we develop an approach that implicitly accounts for periodicity of the crystal lattice. These new techniques are used to study dislocations in tungsten, the main candidate material for plasma-facing armour in future fusion reactors [74,75].

## II. Methodology:

### A. *Manufacture of BCDI strain microscopy samples containing specific defects:*

To introduce glide dislocations, several 500 nm deep Berkovich nano-indents were made into an annealed tungsten crystal with <001> surface normal orientation. Electron channelling contrast imaging (ECCI) was used to identify dislocations for BCDI measurements near these indents. ECCI shows contrast even for small changes in lattice orientation, such as those associated with individual dislocations. Conventional secondary electron imaging would have only shown a flat, featureless surface in this well-polished sample, making it impossible to identify dislocations. More generally, for the manufacture of BCDI samples containing specific microstructural features, any SEM imaging modality that gives contrast for the features of interest, and thus allows them to be identified, can be used.

In ECCI, the sample is placed in an electron channelling condition and a back-scattered electron (BSE) detector is used to record an image (Fig. 1(a)) [76–78]. Here a Zeiss Crossbeam dual beam FIB/FEG SEM with 30 kV acceleration voltage, 10 nA probe current and 8 mm working distance was used. Defect-free regions of the sample appear dark, whilst dislocations appear lighter since their associated lattice distortions locally bring the crystal out of the channelling condition, increasing back-scattered electron yield. As such, the contrast in ECCI is similar to weak-beam dark-field TEM images. By combining ECCI with electron backscatter diffraction (EBSD), the specific lattice planes giving rise to channelling contrast were determined. The associated **q** vector is shown in Fig. 1(a). The two dislocations considered in this study are identified by two grey arrows in Fig. 1(a). They were marked in the Crossbeam microscope by depositing a small amount of carbon on the sample surface.

To produce a sample of suitable size for BCDI measurements and containing the defects of interest, a new FIB sample preparation technique was developed. Manufacture of these samples was carried out on a Zeiss Auriga dual beam FIB/FEG SEM [1]. Initially, using the SEM electron beam in the Auriga instrument to assist deposition (hereafter referred to as e-beam deposition), two ~200 nm wide orthogonal platinum compound (Pt) alignment lines were deposited on the sample, crossing at the site of the defects (Fig. 2(a)). Next a ~4 μm thick Pt cap was deposited on top of the defects, initially using e-beam deposition to protect the sample surface from FIB damage (Fig. 2(b)) (e-beam deposition: 5 keV and very slow scan speed (scan speed 13 on Zeiss instruments). Ga-FIB assisted deposition: 30 kV, 50 pA). Then, following a procedure similar to FIB fabrication of TEM or atom probe tomography (APT) samples [79], FIB milling (30 kV, 16 nA to 1nA) was used to create a ~2 μm wide liftout lamella containing the defects of interest at its centre (Fig. 2(c)). The Pt cap (orange in

---

[1] Two different microscopes were used (Zeiss Crossbeam and Zeiss Auriga). This is because only the Crossbeam instrument was fitted with the BSE detector required for ECCI. Experimental time was more readily available on the Auriga instrument; hence this was used for sample manufacture.

Fig. 2(b) & (c)) was milled into a wedge shape, with the apex aligned with one of the Pt alignment lines (30 KV, 240 pA). The second Pt-alignment line was redeposited over the top of the wedge shape using e-beam Pt deposition. This is important for alignment, as the intersection of the wedge apex and the second Pt alignment line is now directly above the defects of interest. The lamella was then lifted out using a micro-manipulator (yellow in Fig. 2(c) - (f)), and attached to a second, horizontally-mounted needle (purple in Fig. 2(d)). Using this second needle it was turned upside down [80], and then re-attached to the micro-manipulator (Fig. 2(e)). With the Pt wedge facing down, the lamella was then welded to a ~2 μm diameter silicon micro post (supplied by Cameca), using the apex of the wedge and the Pt alignment line to position the defects of interest above the centre of the Si post (Fig. 2(f)). FIB milling (30 kV, 120 pA) was then used to trim down the lamella, leaving a micron-sized sample containing the defects of interest (Fig. 2(g)). Finally, it is vital to remove FIB-induced defects from previous high energy milling steps, which take the form of small dislocation loops and defect clusters [48,63,81]. Whilst the defects caused by high energy FIB milling are confined to a few tens of nanometre thick surface layer, the associated strain fields may extend up to hundreds of nano-metres into the material [39,40,67]. Here, defects from 30 keV FIB milling were removed by low energy FIB polishing (2 keV, 200 pA), ensuring that at least 100 nm of material are removed from all sides of the sample. A scanning electron microscope (SEM) image of the finished sample is shown in Fig. 1(c).

This technique makes it possible to reliably position specific micro-structural features identified in macroscopic samples within a micron-sized volume suitable for BCDI strain microscopy. By initially protecting the top surface and then mounting the sample "upside down" the damage caused by high energy FIB milling operations can be removed from all surfaces of the sample.

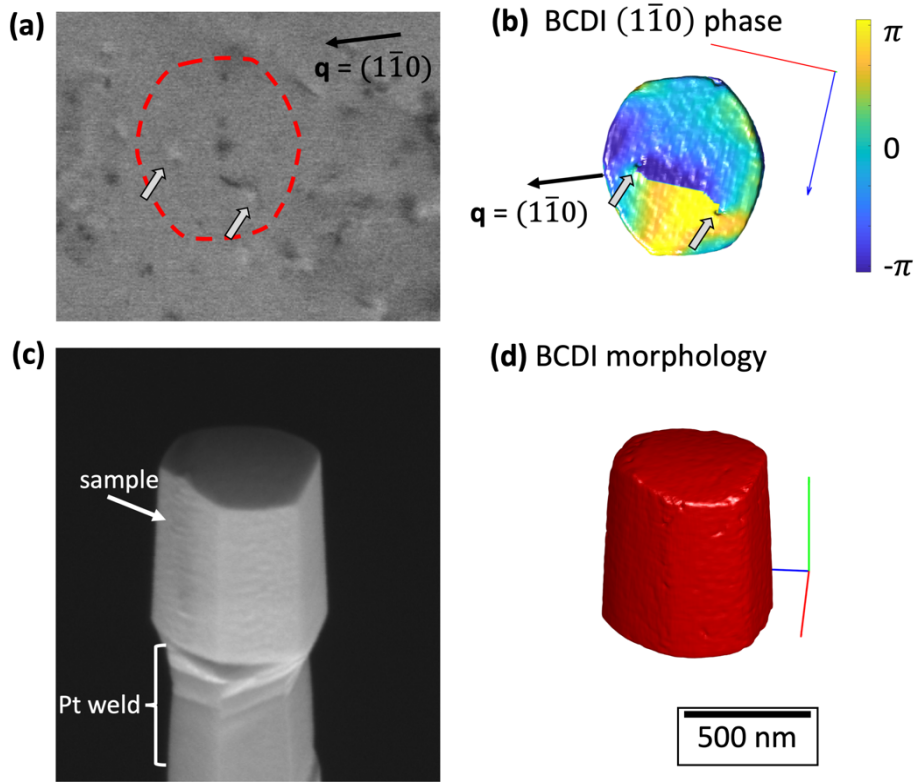

*Figure 1: Selected defects and sample morphology.* (a) ECCI map of the bulk sample surface. Dislocations appears as lighter areas. Two dislocations of interest are marked (grey arrows). The red dashed line shows the outline of the BCDI sample. (b) Sample morphology reconstructed from BCDI measurement of the same $(1\bar{1}0)$ reflection as probed by ECCI. The sample is oriented the same way as in (a) and the two dislocations seen in ECCI are clearly visible (grey arrows). The sample is coloured according to the recovered phase without any phase offset. (c) SEM view of the BCDI sample also showing the Pt weld used for mounting. The surface seen in (a) and (b) is facing down. (d) Rendering of the average sample morphology recovered from BCDI of six different reflections. The same viewpoint and length-scale as for (c) is used. The scalebar in (d) applies to all parts of this figure. Red, green and blue arrows in (b) and (d) show the directions of x, y and z axes respectively and are plotted with a length of 500 nm. The same coordinate system is used throughout this paper.

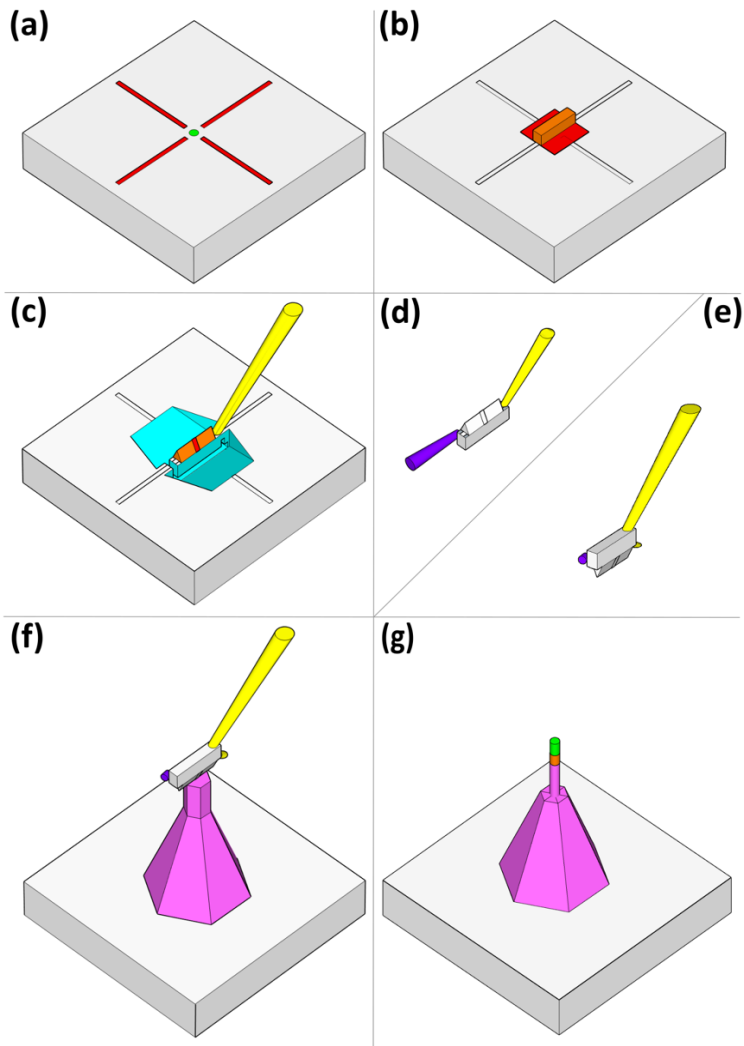

*Figure 2: Strain microscopy sample preparation. (a) Defect of interest (green) is identified on the sample surface and marked using Pt lines (red). (b) Protective Pt is deposited over the defect (red). (c) A lamella containing the defect is milled out using FIB and lifted out using an insitu micro-manipulator (yellow). (d) Using a second needle (purple) the liftout lamella is turned upside down and then reattached to the micro-manipulator (e). (f) The lamella is welded to a silicon post (magenta) with the defect centred above the top of the post. (g) Using FIB the sample is shaped to a micro-crystal containing the defect of interest.*

### B. BCDI measurements and analysis:

Laue micro-diffraction at beamline 34-ID-E at the Advanced Photon Source (APS) was used to determine the crystal orientation of the sample. This served to pre-orient the sample for coherent X-ray diffraction measurements of multiple crystal reflections at beamline 34-ID-C, APS [72]. Coherent X-ray measurements at 34-ID-C used a focussed monochromatic X-ray beam (10 keV photon energy, 1.2 x 1.2 µm$^2$ focus full width at half maximum). While three independent reflections are sufficient for recovery of the full lattice strain tensor, the measurement of additional reflections is beneficial, as it reduces noise in the reconstructed strain tensor. Here we concentrate on {110} crystal reflections, as the {200} reflections could not be reached due to geometrical constraints.

Six {110} crystal reflections were measured: (110), (1$\bar{1}$0), ($\bar{1}$0$\bar{1}$), (10$\bar{1}$), (0$\bar{1}\bar{1}$) and (01$\bar{1}$). For each reflection the crystal was rocked from -0.5° to 0.5° (relative to the reflection centre) in 200 steps with a 1.5 s exposure. This scan was repeated 20 times, aligning the sample to the X-ray beam centre before each scan to compensate for drift. Diffraction patterns were recorded on a Timepix detector (Amsterdam Scientific Instruments) with 256 x 256 pixels and a pixel size of 55 x 55 µm$^2$ positioned at a distance of 1.4 m from the sample. The 20 repeated scans for each reflection were aligned (using a 3D version of the approach proposed by Guizar-Sicairos et al. [82]), summing scans with a cross correlation coefficient greater than 0.987 to return the CXDP used for further analysis of each reflection (number scans summed for each reflection: (110) '15', (1$\bar{1}$0) '16', ($\bar{1}$0$\bar{1}$) '17', (10$\bar{1}$) '12', (0$\bar{1}\bar{1}$) '18' and (01$\bar{1}$) '19'). The resulting diffraction data, which was used for phase retrieval, can be viewed in supplementary movies [83–88] respectively with intensity shown on a log scale.

Well-established phase retrieval approaches [27] were then used to recover the complex-valued electron density from the CXDPs. Phase retrieval was performed iterating between the detector reciprocal space (where the diffraction data is measured) and the detector-conjugated real-space (non-orthogonal coordinate frame conjugated to the detector reciprocal space) [89,90]. Details of the phase retrieval procedure are provided in Appendix A. Finally, the electron density recovered from each reflection was projected back into a common, orthogonal sample coordinate frame [72] with 5 x 5 x 5 nm$^3$ voxel size. Spatial resolution was quantified by differentiating line profiles of electron density amplitude across the object-air interface and fitting these with a Gaussian profile. The average spatial resolution, taken as $2\sigma$ of the fitted Gaussian, is 22 nm.

### III. Results and discussion

*A. Sample morphology and dislocations*

The recovered sample morphology (average of all six reflections, Fig. 1 (d)), is in excellent agreement with an SEM micrograph of the sample recorded from the same view point (Fig. 1(c)). It is worth noting that whilst the SEM image shows the Pt weld attaching the sample to the Si post, this is not seen in the BCDI reconstruction, because it is not part of the coherently scattering domain contributing to the measured reflections. Since the Bragg vector for the ECCI map, $\mathbf{q}_{hkl} = (1\bar{1}0)$, and the relative orientations of the sample in SEM and coherent diffraction measurements are known, the ECCI map (Fig. 1(a)) can be directly compared to the BCDI measurement of the crystal reflection with the same scattering vector. Fig. 1(b) shows the sample morphology recovered from the $(1\bar{1}0)$ reflection, coloured according to the recovered phase. The same two dislocations visible in the ECCI image (marked with grey arrows) can be clearly identified in the BCDI measurement of the $(1\bar{1}0)$ reflection. They appear as two little holes, surrounded by a phase ramp from $-\pi$ to $\pi$, consistent with previous observations of dislocations in single reflection BCDI measurements [30,31,39] . The spacing between the two dislocations agrees very well in ECCI and BCDI. Interestingly the phase jump in the BCDI reconstruction links the two dislocations (Fig. 1(b)), suggesting that they are in fact two ends of the same dislocation line.

The 3D morphology of the sample, recovered from each crystal reflection, is shown in Fig. 3(a), rendered as a semi-transparent iso-surface of electron density amplitude (supplementary movie [91] shows an animated version of Fig. 3(a)). In addition to accurately capturing the finer morphological details (e.g. slight mottling of the surface caused by low energy polishing), channels of reduced electron density crossing the crystal are visible in the reconstructions. Previous simulations of defects in BCDI measurements showed that dislocations appear as pipes of missing electron density [31]. The reason is that large lattice strains near the dislocation core lead to scattered intensity beyond the numerical aperture of the detector, causing an apparent loss of electron density at dislocation cores. By superimposing the electron density recovered from all six measured crystal reflections, five dislocation lines can be segmented (using the Simple Neurite Tracker ImageJ plugin [92]) and are labelled as 1 to 5 in Fig. 3(b) (see also supplementary movie [93]). Closer inspection of Figs. 3(a) and (b) reveals that each dislocation only appears in a subset of reflections. For example, dislocation 2 is seen in the $(1\bar{1}0)$, $(10\bar{1})$ and $(0\bar{1}\bar{1})$ reflections, but not in the other three. This is because only crystal planes that are distorted by a given dislocation will show contrast due to that dislocation. This effect is well known from electron microscopy, where a dislocation only gives rise to contrast in a particular hkl reflection if $\mathbf{q}_{hkl} \cdot \mathbf{b}$ is non-zero, where $\mathbf{b}$ is the dislocation Burgers vector [94]. As $\mathbf{q}_{hkl}$ for each reflection is known and dislocations in tungsten are expected to have either $\frac{1}{2}\langle 111 \rangle$ or $\langle 100 \rangle$ Burgers vector [95], the Burgers vector direction for each dislocation can be determined (see dislocation visibility listed in Table T1 Appendix B). However, $\mathbf{q}_{hkl} \cdot \mathbf{b}$ analysis only allows the Burgers vector direction to be found. To determine the sign of Burgers vector, the associated strain fields must be considered, as discussed below. Using this approach, the Burgers vectors of dislocations 1 and 2 were found to be $\mathbf{b}_1 = \frac{1}{2}[\bar{1}\bar{1}\bar{1}]$ and $\mathbf{b}_2 = \frac{1}{2}[\bar{1}11]$ respectively.

A magnified view of dislocations 1 and 2 is shown in Fig. 3(c) with the Burgers vectors superimposed (black arrows). Interestingly dislocation 1 has a helical shape. Examination of Fig. 3(a) and supplementary movie [91] shows that the helix is visible in every reflection where dislocation 1 features ($(110)$, $(\bar{1}0\bar{1})$ and $(0\bar{1}\bar{1})$ reflections). Helical dislocations are formed when a screw or mixed character dislocation absorbs or emits vacancies, leading to bow-out climb [1]. Considering dislocation line and Burgers vector directions, dislocation 1 can be identified as a right-handed screw dislocation, while the helix is left-handed. This suggests that dislocation 1 helix was formed by the emission of vacancies, or rather by the absorption of interstitials. Dislocation 2 shows a similar, slightly less pronounced helix, which is also consistent with interstitial absorption. Self-interstitials in tungsten delocalise into <111> crowdions and are highly mobile even at cryogenic temperatures [96]. Vacancies, on the other hand, only become mobile above ∼ 600 K [97,98]. As such a dislocation structure driven by interstitial accumulation is expected.

Using $\mathbf{q}_{hkl} \cdot \mathbf{b}$ analysis and considering lattice strains, the Burgers vectors for dislocations 3, 4 and 5 can be determined as $\mathbf{b}_3 = [\bar{1}00]$, $\mathbf{b}_4 = \frac{1}{2}[\bar{1}\bar{1}\bar{1}]$, and $\mathbf{b}_5 = \frac{1}{2}[\bar{1}11]$ respectively. A magnified view shows that dislocations 3, 4 and 5 form a junction in the crystal (Fig. 3(d)). Geometry makes it impossible for dislocations to terminate in the crystal; they must either form a closed loop or a line that emerges at sample surfaces [1]. This means that Burgers vector must be conserved at dislocation junctions (in the present case $\mathbf{b}_3 = \mathbf{b}_4 + \mathbf{b}_5$). This indeed holds true for the Burgers vectors determined above. Dislocation junctions play a central role in controlling the formation of dislocation networks and hence the hardening of crystalline metals. For example, they are responsible for the strongly orientation-dependent strain hardening in bcc metals [70,99] . The fact that BCDI enables such a complete characterisation of the 3D morphology of dislocation junctions is remarkable and will be very useful for validating the substantial body of theoretical predictions of their structure, formation and evolution [70,99–103].

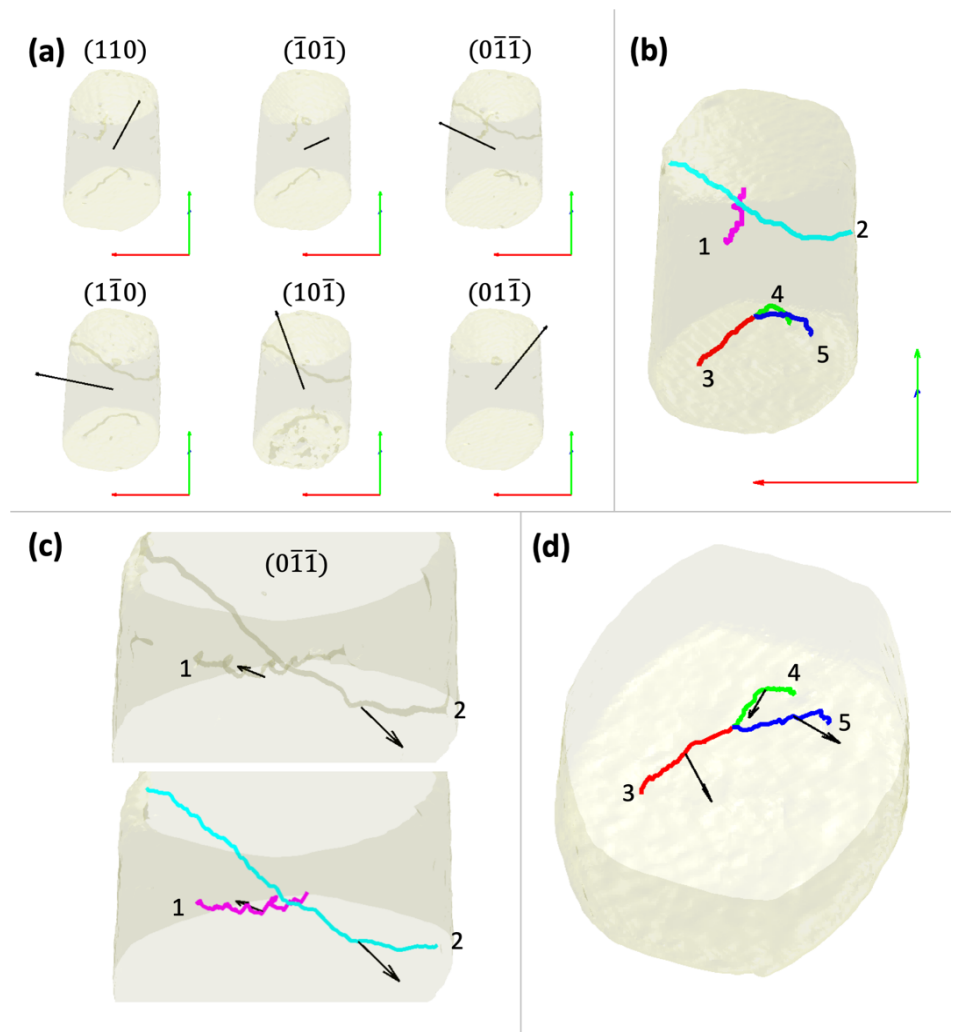

*Figure 3: 3D dislocation structures.* *(a) Semi-transparent 3D rendering of the electron density amplitude recovered from BCDI measurements of six different reflections. Superimposed on each is the scattering vector direction (black arrow). (b) Dislocation lines identified in the sample superimposed on the recovered sample morphology. (c) Detailed view of dislocations 1 and 2, seen as pipes of missing intensity in the electron density recovered from the $(0\bar{1}\bar{1})$ reflection (top). Positions of dislocations 1 and 2 recovered from all reflections (bottom). Superimposed are the Burgers vectors of both dislocations (black arrows). (d) Detailed view of dislocations 3, 4, and 5 showing the morphology of the junction formed by these dislocations. Red, green and blue arrows in (a) and (b) indicate the x, y and z axes directions and are plotted with a length of 500 nm.*

*B. Strains analysis for dislocation-containing samples*

Determination of the full, 3D-resolved lattice strain tensor from BCDI measurement of multiple crystal reflections normally focusses on the reconstruction of the 3D lattice displacement field, $\mathbf{u}(\mathbf{r})$. If at least three independent reflections were measured, this is done by minimising the "squared error"

$$E(\mathbf{r}) = \sum_{hkl}[\mathbf{q}_{hkl} \cdot \mathbf{u}(\mathbf{r}) - \phi_{hkl}(\mathbf{r})]^2, \tag{1}$$

where $\mathbf{q}_{hkl}$ is the Bragg vector of a particular *hkl* reflection, $\phi_{hkl}(\mathbf{r})$ is the phase measured from that reflection and the summation is performed over all measured crystal reflections [44,72]. $E(\mathbf{r})$ is minimised separately for every voxel in the sample. The lattice strain tensor, $\boldsymbol{\varepsilon}(\mathbf{r})$, and rotation tensor, $\boldsymbol{\omega}(\mathbf{r})$, are then obtained by differentiation of the displacement field [104]:

$$\boldsymbol{\varepsilon}(\mathbf{r}) = \frac{1}{2}\{\text{grad } \mathbf{u}(\mathbf{r}) + [\text{grad } \mathbf{u}(\mathbf{r})]^T\}, \tag{2}$$

$$\boldsymbol{\omega}(\mathbf{r}) = \frac{1}{2}\{\text{grad } \mathbf{u}(\mathbf{r}) - [\text{grad } \mathbf{u}(\mathbf{r})]^T\}. \tag{3}$$

This approach works well if the phase variation is smooth. Problems arise if there are jumps in the phase. Even for a smoothly varying displacement field these jumps may arise due to wrapping of the phase if the lattice displacement magnitude in the direction of $\mathbf{q}_{hkl}$ is greater than $\pi/|\mathbf{q}_{hkl}|$, since the phase is defined between $-\pi$ and $\pi$. Phase unwrapping algorithms [73] can be used to unwrap these jumps, after which equation (1) can be used to reconstruct the lattice displacement field [72]. Phase jumps due to crystal defects present more of a challenge. They cannot be removed by phase unwrapping as they do not traverse the whole crystal, but end at the dislocation lines (see Fig. 1(b)). Physically, the phase jump associated with a dislocation corresponds to the plastic deformation mediated by that dislocation, and its magnitude is given by $\Delta\phi_{hkl} = \mathbf{b} \cdot \mathbf{q}_{hkl}$ [39]. Since lattice planes are indistinguishable, the position of the phase jump due to a specific dislocation is not uniquely defined. It can be moved around simply by adding a phase offset, i.e. selecting a different zero-phase reference, as illustrated in Fig. 4(a). As a result, the phase jumps due to a particular dislocation are generally not in exactly the same position in BCDI measurements of different reflections containing that dislocation. This complicates determination of the correct displacement tensor. Furthermore, numerical differentiation of a discontinuous displacement tensor, required for the determination of lattice strain, will lead to incorrect, large strains at discontinuities unless periodicity of the crystal lattice is accounted for. This is not straightforward to do in a general way. Supplementary figure [105] illustrates how displacement-field-based analysis fails in the present sample, leading to large, spurious strains.

We propose a new approach for computing lattice strain from multi-reflection BCDI measurements. Since $\phi_{hkl}(\mathbf{r}) = \mathbf{q}_{hkl} \cdot \mathbf{u}(\mathbf{r})$, the spatial derivatives of $\phi_{hkl}(\mathbf{r})$ are:

$$\frac{\partial \phi_{hkl}(\mathbf{r})}{\partial i} = \mathbf{q}_{hkl} \cdot \frac{\partial \mathbf{u}(\mathbf{r})}{\partial i}, \tag{4}$$

where $i$ refers to the spatial $x$, $y$ or $z$ coordinate. An optimisation problem, similar to equation (1), can then be formulated to find the spatial derivatives of the displacement field from the phase gradients by minimising the "squared error"

$$E(\mathbf{r})_i = \sum_{hkl,i} \left[ \mathbf{q}_{hkl} \cdot \frac{\partial \mathbf{u}(\mathbf{r})}{\partial i} - \frac{\partial \phi_{hkl}(\mathbf{r})}{\partial i} \right]^2, \qquad (5)$$

where *hkl* refers to the measured reflections and the summation is carried out over all measured reflections. By performing this optimisation, all components of the displacement gradient $\boldsymbol{\beta}(\mathbf{r}) = \text{grad}\,\mathbf{u}(\mathbf{r})$ can be found, and hence the lattice strain and rotation tensor evaluated (equations 2 and 3).

This approach dramatically simplifies the computation of $\boldsymbol{\varepsilon}(\mathbf{r})$ and $\boldsymbol{\omega}(\mathbf{r})$ since the phase gradients for each reflection can be readily computed. Guizar-Sicairos et al [106] proposed a convenient method for the numerical calculation of phase gradients in the presence of phase jumps, provided these jumps are "sharp". A complication in the present case is that phase derivatives are required in an orthogonal $x$, $y$, $z$ sample coordinate frame common to all reflections. The complex electron density associated with each reflection, however, is retrieved in non-orthogonal detector conjugated space (see section II.B above). Mapping of the data from the detector conjugated space to the orthogonal sample coordinate frame necessarily involves an interpolation step. This can lead to a spreading-out of phase jumps over two or more pixels (depending on interpolation spacing), meaning that phase jumps are no longer "sharp" and will give rise to errors when phase derivatives are calculated.

Here a modified approach is used: For the phase recovered from each reflection two additional phase-shifted copies are generated in detector conjugated space by adding phase offsets of $-\frac{\pi}{2}$ and $\frac{\pi}{2}$ to the reconstruction. Any voxels with a phase outside the range $-\pi$ to $\pi$ are returned to this range by adding or subtracting $2\pi$. This implicitly imposes continuity and periodicity of the crystal lattice and shifts phase jumps associated with the dislocation to different positions (Fig. 4(a)). Carrying out this operation in the detector conjugated space used for phase retrieval guarantees "sharp" phase jumps. Next the original and shifted phases are transformed to the orthogonal sample coordinate frame, common to all reflections, and the spatial phase derivatives are computed using numerical differentiation (Fig. 4(b)). These phase gradients show large spurious values associated with the phase jumps. However, the spurious values are located in different positions for each phase offset. By selecting the phase gradient with the smallest magnitude for each voxel, the corrected phase derivatives can be found (Fig. 4(c)). This approach allows a straightforward calculation of the phase gradients required for the computation of the strain and rotation tensors, which are found by applying equation 5 in the common, orthogonal coordinate frame shared by all reflections.

The reconstructed components of the lattice strain and rotation tensors on a virtual section through the sample are shown in Fig. 5(b) (see supplementary movie [107] for the strains reconstructed throughout the volume). The local strain and rotation fields associated with dislocations 2,4 and 5, which intersect the plotting plane (Fig. 5(a) and supplementary movie [108]), can be clearly identified. A comparison with supplementary figure [105] illustrates that the new approach removes the large, spurious strains recovered using conventional displacement-field-based analysis.

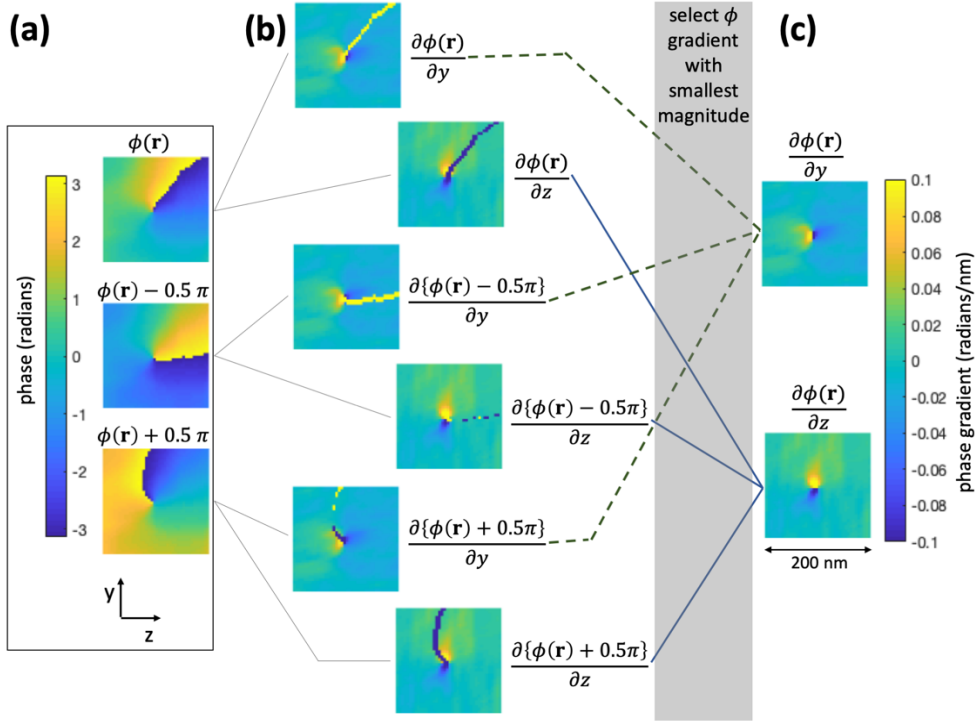

*Figure 4: Computation of phase gradients near a dislocation.* (a) Reconstructed phase of the complex electron density with phase offsets of 0, $-\frac{\pi}{2}$ and $\frac{\pi}{2}$. The phase is remapped into the $-\pi$ to $\pi$ range, changing the position of the phase jump associated with the dislocation. (b) Phase gradients in y and z directions for all three phase offsets. In each map a line of large spurious phase gradient, due to the phase jump, is clearly visible. (c) Filtered phase gradients in y and z directions where for each pixel the phase gradient with the lowest magnitude is used. Plots in (b) and (c) are plotted on the same colour scale and all plots measure 200 x 200 nm². The phase signal is from dislocation 2 in the $(10\bar{1})$ reflection, shown in the y-z plane at position x = 0 nm.

C. 3D strain fields – measured and predicted:

To provide a direct quantitative comparison for the complicated lattices strain fields measured experimentally, a 3D dislocation model was constructed. The 5 dislocations identified in the sample were discretised into lines of points (Fig. 5(a)). These points were linked by dislocation segments to which the corresponding Burgers vector, found by $\mathbf{q}_{hkl} \cdot \mathbf{b}$ analysis, was assigned. Each dislocation line was then linked to a remote closure point outside the sample to form a dislocation triangle. The lattice strain and rotation fields caused by dislocation triangles were determined using numerical differentiation of the solution developed by Barnett for the displacement field of a triangular dislocation loop in an infinite, elastically isotropic medium [109,110]. This is appropriate since tungsten is almost perfectly elastically isotropic [111–113]. By superimposing the strain and rotation fields of all dislocation triangles, the overall distortion fields in the sample due to the dislocation lines can be predicted [114]. Where dislocations emerge at sample surfaces, the dislocation line was extended a further 1 μm outside the sample, normal to the sample surface. The effect of surface relaxation, due to traction free boundary conditions on the sample surface, was not accounted for. This is acceptable since material near the sample surface will probably be affected by spurious strains due to residual FIB damage [39,40], obviating the need to account for surface relaxation, the effects of which diminishes beyond depths of a few 10s of nm [115].

The predicted strains and lattice rotations, plotted on the same virtual slice through the crystal as those measured by BCDI (Fig. 5(b)) are shown in Fig. 5(c) (see supplementary movie [116] for strains predicted throughout the sample volume). The agreement is striking. Not only are the magnitudes of lattice strains near dislocations captured correctly, even subtleties, such as the overlapping strain and rotation fields associated with dislocations 4 and 5 are correctly captured. This excellent agreement is not only obtained for the cross-section shown in Fig. 5, but throughout the crystal. An important point here is that the strains only match if both Burgers vector direction (determined from $\mathbf{q}_{hkl}.\mathbf{b}$ contrast) and sign are correct. Reversing the sign of Burgers vector will reverse the sign of the strain fields. Hence, by comparing the simulated and measured strain fields, the full Burgers vector of each dislocation was unambiguously determined. The excellent agreement of the lattice strains can be verified by viewing supplementary movies [107] and [116] that show measured and predicted distortions respectively on y-z slices through the sample for different x-positions.

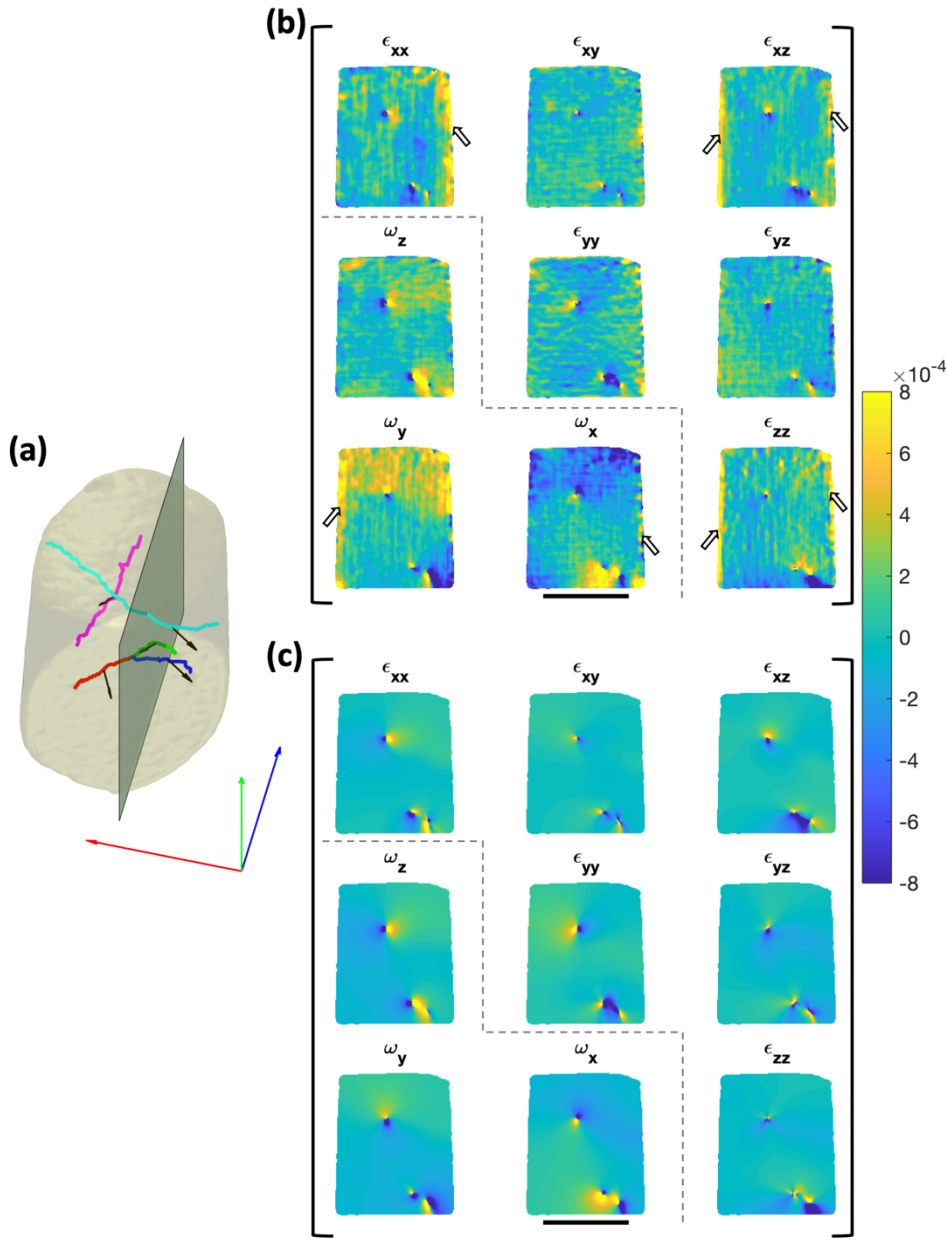

*Figure 5: Lattice strains in the sample: Measurement and Simulation.* (a) Rendering of the 3D sample morphology showing dislocation lines and Burgers vectors (black arrows). Superimposed in green is the y-z plane on which strains in (b) and (c) are plotted. The red, green and blue arrows indicate the direction of x, y and z axes respectively and are plotted with a length of 500 nm. (b) Lattice strains measured using multi-reflection BCDI. Shown are the six components of the lattice strain tensor (upper triangle and diagonal), and the three components of the lattice rotation tensor (lower triangle). The strain fields of dislocations 2, 4 and 5, which intersect the plotting plane, can be readily identified. Hollow arrows point to the surface strain associated with residual FIB damage. (c) Lattice strain fields predicted using a 3D model of the dislocation structures in the sample plotted on the same plane as (c). Lattice strains and rotations (in radians) in (b) and (c) are plotted on the same colour scale and at the same magnification. The scalebar in (b) and (c) is 500 nm long.

To quantitatively assess agreement of experimentally-measured and predicted strain fields, the mean difference strain, $\overline{\Delta\varepsilon}$, and mean rotation difference, $\overline{\Delta\omega}$, can be considered:

$$\overline{\Delta\varepsilon} = \frac{1}{V}\int_V |\varepsilon_{exp} - \varepsilon_{sim}|dV \qquad \text{and} \qquad \overline{\Delta\omega} = \frac{1}{V}\int_V |\omega_{exp} - \omega_{sim}|dV, \qquad (6)$$

where $\varepsilon_{exp}$ and $\varepsilon_{sim}$ are the strains, and $\omega_{exp}$ and $\omega_{sim}$ are the lattice rotations measured experimentally and predicted by simulations, respectively. By only considering material more than 30 nm away from dislocation lines or free surfaces, the agreement for low strain material volumes can be probed. In this case

$$\overline{\Delta\varepsilon} = \begin{bmatrix} 1.9 & 1.2 & 1.2 \\ 1.2 & 1.4 & 1.3 \\ 1.2 & 1.3 & 1.6 \end{bmatrix} \times 10^{-4} \qquad \text{and} \qquad \overline{\Delta\omega} = \begin{bmatrix} 1.6 \\ 2.4 \\ 1.7 \end{bmatrix} \times 10^{-4}.$$

This suggests a strain uncertainty in the measurement of ~2 x $10^{-4}$ for weakly strained volumes, consistent with our previous estimate of strain uncertainty in MBCDI measurements [40].

When considering only highly strained material (here taken as being within 30 nm of a dislocation line), $\overline{\Delta\varepsilon}$ and $\overline{\Delta\omega}$ increase slightly to:

$$\overline{\Delta\varepsilon} = \begin{bmatrix} 4.2 & 2.9 & 3.6 \\ 2.9 & 4.6 & 3.8 \\ 3.6 & 3.8 & 4.8 \end{bmatrix} \times 10^{-4} \qquad \text{and} \qquad \overline{\Delta\omega} = \begin{bmatrix} 4.7 \\ 4.5 \\ 4.3 \end{bmatrix} \times 10^{-4}.$$

An approach frequently used in the dislocation community to visualise dislocation strain fields is to consider strain variation along a circular path around a given dislocation [30,117,118]. Here we consider strain variation along circular paths with 30 nm radius around dislocations 2, 4 and 5 that cross the virtual section through the crystal in Fig. 5, as shown in Fig. 6(a). The variation of all six lattice strain and three lattice rotation components, plotted as a function of angle around the circular path, is shown in Fig. 6(b), (c) and (d) for dislocations 2, 5 and 4 respectively. The strain and rotation profiles measured experimentally and predicted using the dislocation statics model are plotted in red and blue respectively. They agree very well, with the experimental measurement even capturing rather subtle features, for example the "double hump" structure of $\varepsilon_{yy}$ for dislocation 2 (Fig. 6 (b)). The mean difference between the measured and predicted lattice strains and rotations for all strain profiles in Fig. 6 is ~2 x $10^{-4}$. These results show that MBCDI can be used to reliably map out the full, 3D resolved strain and lattice rotation fields associated with tortuous dislocations.

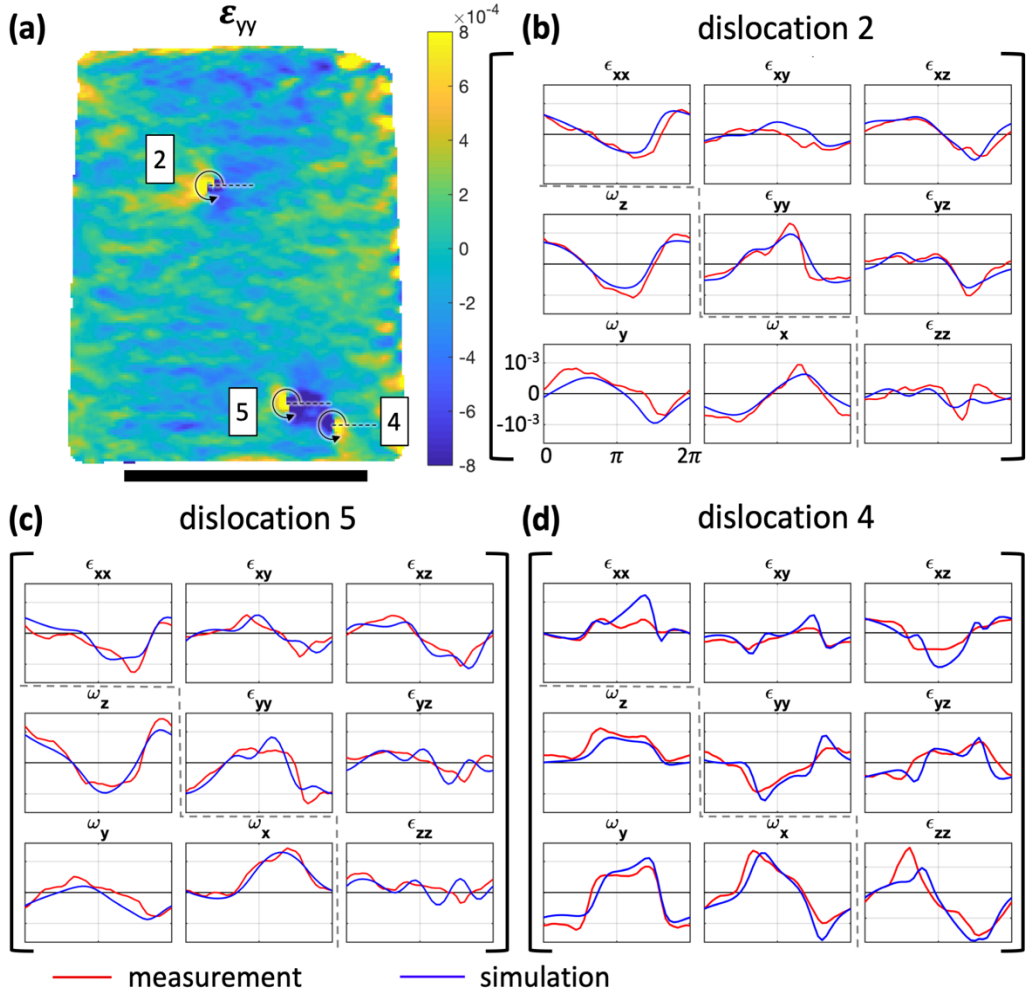

*Figure 6: Quantitative comparison of predicted and measured lattice strain fields near dislocations.* (a) Plot of the measured $\varepsilon_{yy}$ lattice strain field on the same virtual slice through the sample considered in Fig. 5. Superimposed are labels identifying the three dislocations crossing this slice (dislocations 2, 4 and 5). The scalebar in (a) is 500 nm long. (b), (c) and (d) show lattice strain and rotation along a circular path around each dislocation at a radial distance of 30 nm. The circular paths considered are plotted in (a), with the dashed line indicating the angular zero position and the circular arrow indicating the positive angle direction. For each dislocation the six components of the lattice strain tensor and three components of the lattice rotation tensor are plotted. Measured profiles are plotted in red, and predictions from simulations are plotted as blue lines. The y-axis shows strain or rotation (in radians), with the horizontal grid lines corresponding to $10^{-3}$ and $-10^{-3}$ respectively as labelled in the $\omega_y$ component in (b). The x-axis shows the angular position along the circular path in radians from 0 to $2\pi$, as indicated for the $\omega_y$ component in (b). The convention used for plotting of x- and y-axes is the same for all plots in (b), (c) and (d).

D. FIB damage:

A key concern in FIB preparation of microscopy samples is the effect of FIB-induced damage. Previously we showed that even low dose gallium ion exposure causes large lattice strains that can extend 100 nm or more into the sample [39,40]. Several approaches have been proposed for reducing/removing FIB damage in TEM sample preparation. The most attractive is low energy ion-milling, which uses acceleration voltages below 5 kV to remove a shallow surface layer containing the damage introduced by previous high energy milling steps [48,119]. In the tungsten liftout sample, 2 kV milling was used to polish off damage. The reconstructed strain maps show increased lattice strains at the sides of the sample (hollow arrows in Fig. 5(b)). These are not predicted by the dislocation simulation and are attributed to residual FIB-induced defects from the final 2 kV polishing step (see Fig. 5(b) $\varepsilon_{xx}$, $\varepsilon_{xz}$ and $\varepsilon_{zz}$). The apparent thickness of the FIB-induced strained layer in Fig. 5(b) is ~25 nm. Since this is close to the spatial resolution of our measurement (~22 nm), the actual strained layer thickness will be less, on the order of 10 to 15 nm. This is consistent with our previous observation of a ~ 20 nm thick strained layer after 5 kV gallium polishing in gold [40]. It is encouraging that FIB damage in the present tungsten sample could be successfully removed although the sample was exposed to extensive high energy FIB milling during preparation. Transmission electron microscopy has shown that FIB damage in metals and alloys takes a surprisingly similar form, irrespective of the exact elements under consideration [63,120]. As such, the sample preparation approach developed here should be applicable to all metals and alloys, and provide a general tool for the extraction of strain microscopy samples.

## IV. Concluding remarks

It is interesting to compare the dislocation structure analysis with the state of the art in TEM, where substantial scientific effort has been dedicated to the reconstruction of 3D dislocation structures [121–128]. Diffraction contrast TEM has been successfully used for the 3D reconstruction of dislocations visible for a specific $\mathbf{q}_{hkl}$ vector [121,125,128]. To ensure all dislocations in the sample are captured, repeated measurements of three or more $\mathbf{q}_{hkl}$ vectors would be required. Importantly these measurements would not provide any information about the 3D lattice strains associated with dislocations. For very small samples, less than 10 nm in size, 3D-resolved lattice strain mapping with atomic resolution has been demonstrated [122,123]. The BCDI measurements presented here allow detailed analysis of dislocation structures in micron-sized samples extracted from bulk material. By considering multiple Bragg reflections, all dislocations are probed, and their 3D morphology and Burgers vector can be extracted. The measured dislocation strain fields in tungsten are in excellent agreement with predictions from an elastically isotropic model of the dislocation structure. This provides confidence for the application of multi-reflection BCDI strain microscopy to more complex scenarios, such as dislocations in elastically anisotropic crystals, or interactions of defects with precipitates and second phases. Because the FIB preparation approach is highly site specific, it makes it possible to reliably place specific microstructural features of interest within BCDI samples. We anticipate that these developments will broaden the applicability of BCDI to previously inaccessible metallic materials, with applications across material science, condensed matter physics, nano-science and chemistry.

Diffraction data, reconstructions, analysis codes and dislocation simulations presented in this paper can be downloaded from https://doi.org/10.5287/bodleian:0oqB6K7gv or from

https://github.com/Hofmann-Group/Nano-scale-imaging-of-the-full-strain-tensor-of-specific-dislocations-extracted-from-a-bulk-sample.

**Acknowledgements**
The authors would like to thank Sinead Hofmann, Edmund Tarleton, Sergei Dudarev, Ross Harder and Andrew London for insightful discussions; Jesse Clark and Mathew Cherukara for providing the code used for phase retrieval; and Michael Rieth for providing the tungsten material. FH and NWP acknowledge funding from the European Research Council (ERC) under the European Union's Horizon 2020 research and innovation programme (grant agreement No 714697). SD acknowledges support from The Leverhulme Trust under grant RPG-2016-190. JOD was supported by EPSRC Grant EP/P005101/1. The authors acknowledge use of characterisation facilities within the David Cockayne Centre for Electron Microscopy, Department of Materials, University of Oxford and use of the University of Oxford Advanced Research Computing (ARC) facility http://dx.doi.org/10.5281/zenodo.22558. The Zeiss Crossbeam FIB/SEM used in this work was supported by EPSRC through the Strategic Equipment Fund, grant EP/N010868/1. Diffraction experiments used the Advanced Photon Source, a US Department of Energy (DOE) Office of Science User Facility operated for the DOE Office of Science by Argonne National Laboratory under Contract No. DE-AC02-06CH11357.

**Appendix A: Phase retrieval**
Well established phase retrieval approaches for BCDI data [27,31,39] were used to recover the complex electron density from the measured diffraction patterns of the different crystal reflections. The phasing was carried out in four stages, using the output from the previous stage to seed the next phasing stage.
**Stage 1:** Diffraction patterns were cropped to a size of 146 x 146 x 172 pixels and the reconstruction was seeded with a random phase guess. A guided phasing approach with 40 individuals and 4 generations was used [31]. Low resolution data was phased in the first and second generation and full resolution data from generation three onwards. A combination of error reduction (ER) and hybrid input output (HIO with $\beta = 0.9$) iterations was used [129]. For each generation a pattern of 20 ER and 180 HIO iterations was repeated six times followed by 20 ER iterations. The returned object was the average taken over the final 20 ER iterations. A sharpness criterion was used to determine the best reconstruction as this was previously identified as the most suitable metric for crystals containing defects [31,69]. The real-space support was periodically updated using the shrinkwrap algorithm [130].
**Stage 2:** The result from stage 1 was used to seed this reconstruction. All parameters were kept the same as for stage 1. In addition partial coherence effects were corrected for using the approach proposed by Clark et al. [27]. This accounts for partial coherence in both longitudinal and transverse directions. Here a general coherence function was used, determined by Richardson-Lucy deconvolution.
**Stage 3:** The full 256 x 256 x 256 pixels CXDP for each reflection (zero-padded to 256 pixels in the 3rd direction) was used as input and the reconstruction was seeded with the re-sized output object from stage 2. Partial coherence was included in the refinement, but no guided phasing was used. A pattern of 20 ER and 180 HIO iterations was repeated 15 times, followed by 1000 iterations of ER. The returned object was the average of the final 50 ER iterations.

**Stage 4:** The previous stage of phasing (stage3) was repeated using the output from stage 3 to seed the reconstruction.

Next, three versions of the electron density reconstructed from each crystal reflection were generated with phase offsets of 0, $-\frac{\pi}{2}$ and $\frac{\pi}{2}$. Voxels with a phase outside the range $-\pi$ to $\pi$ were returned to this range by adding or subtracting $2\pi$. Finally, all reconstructions were transformed from the detector conjugated space used for phase retrieval to an orthogonal sample space with a voxel size of 5 x 5 x 5 nm$^3$. The twin ambiguity of BCDI was overcome by comparing the morphology retrieved from BCDI to SEM images of the sample recorded in the same orientation (Fig. 1). The overall reconstructed object is the average of the electron density amplitudes recovered from the six measured reflections.

**Appendix B: Visibility of dislocations in different reflections**

| q \ b | (110) | (1$\bar{1}$0) | ($\bar{1}$0$\bar{1}$) | (10$\bar{1}$) | (0$\bar{1}\bar{1}$) | (01$\bar{1}$) |
|---|---|---|---|---|---|---|
| $\frac{1}{2}[111]$ | visible | | visible | | visible | |
| $\frac{1}{2}[\bar{1}11]$ | | visible | | visible | visible | |
| $\frac{1}{2}[1\bar{1}1]$ | | visible | visible | | | visible |
| $\frac{1}{2}[11\bar{1}]$ | visible | | | visible | | visible |
| [100] | visible | visible | visible | visible | | |
| [010] | visible | visible | | | visible | visible |
| [001] | | | visible | visible | visible | visible |

***Table T1***: *Visibility of dislocations with Burgers vector* **b** *in crystal reflections with scattering vector* **q**.

*predicted by a linear elastic model of the 3D dislocation structure. The three components of the lattice rotation tensor predicted by this model are also shown (lower triangle). The positioning of slices, colour scale and length-scale are the same as for the experimentally measured lattice distortions shown in supplementary movie [107]. Horizontal and vertical axes show length in nm..*